\documentclass[preprint,12pt,authoryear]{elsarticle}

\usepackage{amssymb}
\usepackage{amsmath}
\usepackage{graphicx}%
\usepackage{multirow}%
\usepackage{amsmath,amssymb,amsfonts}%
\usepackage{amsthm}%
\usepackage{mathrsfs}%
\usepackage[title]{appendix}%
\usepackage{xcolor}%
\usepackage{textcomp}%
\usepackage{manyfoot}%
\usepackage{booktabs}%
\usepackage{algorithm}%
\usepackage{algorithmicx}%
\usepackage{algpseudocode}%
\usepackage{listings}%
\usepackage{subcaption}

\usepackage{graphicx}
\usepackage{booktabs}
\usepackage{tabularx}
\usepackage{makecell} 
\usepackage{multirow}

\usepackage{listings}
\usepackage{subcaption}

\usepackage{amssymb}
\usepackage{amsmath}

\usepackage{tabularx}
\usepackage{xcolor}
\usepackage{url}
\usepackage{booktabs}
\usepackage{graphicx} 
\usepackage[most]{tcolorbox}
\usepackage{lipsum}
\usepackage{tabularx}
\usepackage{booktabs}
\usepackage{geometry}
\usepackage{longtable}
\usepackage{array}
\usepackage{caption}
\usepackage{graphicx}
\usepackage{subcaption}

\usepackage{float}
\usepackage{adjustbox}   

\journal{Nuclear Physics B}

\begin{document}

\begin{frontmatter}



\title{ELEVATE: Designing Human-Centered GenAI Virtual Tutors for Scalable and Inclusive Education}

\author[unimc]{Lorenzo Stacchio\corref{cor1}}
\ead{lorenzo.stacchio@unimc.it}

\author[ucb]{Michele Giordano}
\ead{giordano.michele@gmail.com}

\author[iit]{Daniele Berardini}
\ead{daniele.berardini@iit.it}

\author[univpm]{Primo Zingaretti}
\ead{p.zingaretti@staff.univpm.it}

\author[unimc]{Emanuele Frontoni}
\ead{emanuele.frontoni@unimc.it}

\cortext[cor1]{Corresponding author.}

\affiliation[unimc]{
    organization={University of Macerata},
    addressline={Department of Political Sciences, Communication and International Relations},
    city={Macerata},
    postcode={62100},
    country={Italy}
}

\affiliation[ucb]{
    organization={eCampus University, Facoltà di Ingegneria, Italy},
    city={Como},
    country={Italy}
}

\affiliation[iit]{
    organization={Italian Institute of Technology},
    city={Genoa},
    country={Italy}
}

\affiliation[univpm]{
    organization={Universit\`a Politecnica delle Marche},
    city={Ancona},
    postcode={60131},
    country={Italy}
}

\begin{abstract}
The advent of Generative Artificial Intelligence (GenAI), and in particular Large
Language Models (LLMs), is reshaping educational practice, while intensifying
ethical debate about its adoption. To date, the dominant paradigm remains
cloud-based and text-only chatbot: a centralized service that offers limited pedagogical
control, weak transparency over knowledge sources, and non-trivial risks for privacy
and regulatory compliance. This model also presumes continuous connectivity and recurring
API costs, creating structural barriers for many institutions, reinforcing existing digital divides.
At the same time, educational interaction with LLM can benefit from multimodal cues and embodied presence, requiring interfaces that move beyond text-only tutoring. In this work, we propose ELEVATE (Efficient LLM Education with Virtual Avatar Teaching Engine), a framework to develop efficient GenAI-driven avatar tutors governed by epistemic infrastructures.
ELEVATE integrates LLM-driven dialogue with embodied 3D avatars for multimodal interaction and adopts a local-first execution model enabling deployment on consumer-grade hardware. The framework formalizes a three-stratum design
that separates (i) a student-facing virtual avatar interaction layer, (ii) a local GenAI execution and multimodal synthesis core, and (iii) a teacher-facing governance layer. We implemented and evaluated a working prototype deployed in a real-world educational curriculum. The system runs on standard PCs and smartphones, and we provide system-level performance evidence to show responsive interaction under realistic hardware constraints. Finally, we discuss sociotechnical and pedagogical implications for responsible adoption, positioning ELEVATE as a scalable pathway for privacy-preserving and inclusive GenAI tutoring across heterogeneous school environments.
\end{abstract}

\begin{graphicalabstract}
\includegraphics[width=\textwidth]{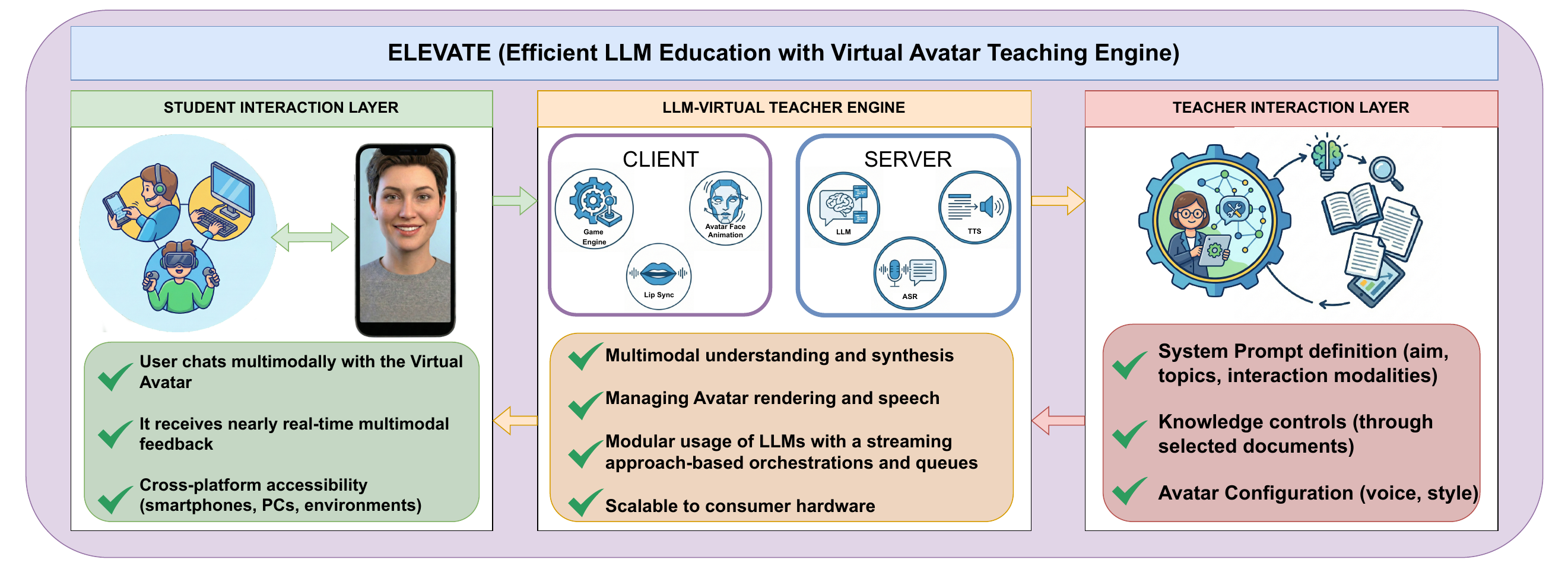}
\end{graphicalabstract}

\begin{highlights}
\item Proposes ELEVATE, a local-first GenAI avatar tutor framework for schools
\item Three-layer architecture separates student interaction, inference, and teacher governance
\item Runs on consumer PCs and smartphones with streaming, near real-time responses
\item Reports latency and perceived-latency metrics; discusses ethics, privacy, and inclusion
\end{highlights}


\end{frontmatter}



\section{Introduction}

The advent of Generative Artificial Intelligence (GenAI) has rapidly transformed educational discourse and practice, offering unprecedented possibilities while simultaneously intensifying debate among educators, institutions, and policymakers. In particular, Large Language Models (LLMs), now adopted by hundreds of millions of users worldwide, are increasingly used by a wide range of educational actors, from students to teachers and administrators~\citep{zhu2024embracing,ma2025systematically,park2025systematic}. Their diffusion has prompted schools and universities to reimagine pedagogical practices and institutional policies, as LLMs promise highly personalized learning experiences, automated content generation, and real-time tutoring adapted to individual students’ pace and interests~\citep{wang2024large,xu2024large,yigci2025large}.
\\
At the same time, the educational adoption of GenAI has raised significant concerns regarding academic integrity, the evolving role of teachers, and the ethical boundaries of human-machine collaboration~\citep{yan2024practical}. These concerns have been explicitly acknowledged by international organizations such as UNESCO, which has called for a human-centered approach to educational AI, one that protects learner privacy, upholds inclusion and equity, and positions AI systems as augmenting rather than replacing human educators~\citep{UNESCO2023GenAI}. 
Such calls underscore that the challenge of GenAI in education is not merely technical, but deeply epistemic and institutional.
\\
Despite these concerns, the dominant paradigm of LLM adoption in education remains the cloud-based, text-only, and non-compliant chatbot~\citep{yigci2025large,pozdniakov2024large}: 
a centralized service that delivers fluent responses but offers limited expressiveness, minimal pedagogical control, and little transparency regarding knowledge sources or interaction boundaries. 

Indeed, State-of-the-art LLMs are typically deployed as centralized cloud services, requiring substantial computational resources and continuous internet connectivity. This raises significant challenges for scalability, accessibility, and sustainability: many educational institutions (particularly those in under-resourced or rural contexts) cannot rely on high-bandwidth connections or afford recurring API costs for large-scale student use. Moreover, centralized architectures raise concerns about data privacy and regulatory compliance, as sensitive student data may be transmitted off-site, complicating adherence to frameworks such as the European GDPR and the forthcoming AI Act~\citep{EU2024AIAct}. Without alternative deployment models, GenAI risks becoming available primarily to well-funded, well-connected classrooms, thereby exacerbating existing digital divides.

Moreover, text-only chatbots and question-answering assistants represent the prevailing form of AI tutors, yet they lack the expressive richness characteristic of human teaching. Prior work suggests that the integration of non-verbal cues (e.g., tone of voice, facial expressions, and embodied presence) can improve clarity, engagement, and resonance in AI-student interactions~\citep{wang2024large,xu2024large,pozdniakov2024large}. Recent advances in human-computer interaction further indicate that moving beyond text-centric interfaces toward embodied, multimodal virtual humans may substantially enhance educational experiences~\citep{john2024llm} while supporting more inclusive educational interactions~\citep{Kacorri2020SAIL,Maresca2025LLMsVisualImpairments,zhao2024embodied}. Indeed speaking and gesturing avatars can benefit learners who struggle with text-based communication (e.g., young children or students with reading difficulties) and can offer accessible interaction modalities for individuals with impairments through voice output, lip-reading support, or sign-language–capable avatars~\citep{Kacorri2020SAIL,Maresca2025LLMsVisualImpairments,park2025systematic}. 

These developments point to a central challenge: realizing the pedagogical potential of GenAI in education requires not only richer interaction modalities but also infrastructural models that align with institutional responsibility, data sovereignty, and ethical governance. In this respect, a growing body of work has argued for local-first and on-premises deployment strategies, especially in sensitive domains such as education, where privacy-by-design and accountability are strict requirements~\citep{stacchio2024xrai,du2024integrating,giordanoLLM2025}. Such approach ensures that student data and interactions remain private to the school environment, directly addressing the European data sovereignty concerns~\citep{UNESCO2023GenAI,EU2024AIAct} and reducing dependency on external providers.
In light of these considerations, and without aiming to evaluate learning outcomes directly, we adopt a system-level perspective on how GenAI-driven Avatar tutoring infrastructures can be designed to align with institutional responsibility, pedagogical governance, and inclusive educational practice.
\\
We here introduce ELEVATE (Efficient LLM Education with Virtual Avatar Teaching Engine), a framework that reconceptualizes GenAI tutors with a scalable and inclusive approach, founded on institutionally governed epistemic infrastructures. ELEVATE integrates LLM-driven dialogue with 3D embodied virtual avatars to support multimodal interaction, while adopting a local-first execution model that enables deployment on consumer-grade hardware. Relevantly, the framework introduces a modular three-stratum design that explicitly separates (i) a Virtual Avatar student-facing interaction, (ii) GenAI execution and synthesis, and (iii) teacher-facing governance. This separation allows to control pedagogical intent, content boundaries, and ethical constraints, that can be explicitly encoded and enforced at the system level only by authorized actors (e.g., teachers). To demonstrate the practicality of this approach, we implement a working prototype of an LLM-driven virtual tutor deployed in a real-world educational context, aligned with curriculum requirements. The prototype runs on affordable devices typical of public school settings, including standard PCs and smartphones, and achieves near real-time end-to-end interaction without reliance on cloud-based GPUs. 
On top of such results, we discuss, with an interdisciplinary perspective, on the role of ELEVATE in education, highlighting how efficiency-oriented and locally deployable architectures can support privacy, accessibility, and inclusive pedagogical practices while defining novel challenges for the educational domain.
\\
Our key contributions can be summarized as follows:
\begin{itemize}
    \item We introduced ELEVATE, a general-purpose framework for multimodal GenAI tutoring, combining LLM with embodied 3D avatars to enable more human-like, engaging interactions across diverse educational levels and subjects.
    \item We detailed a system architecture that runs on low-cost PCs and smartphones entirely on-site. This design ensures data privacy and compliance (no student data leaves the school), reduces latency and bandwidth dependence, and makes advanced AI accessible to resource-constrained schools (prioritizing efficiency, local deployment, and accessibility).
    \item We demonstrated its feasibility through a real-world implemented prototype, with a concrete case study, which illustrates that our approach can achieve real-time performance on modest hardware, and it showcases the potential for pedagogical transformation.
    \item We provided a thorough discussion on the responsible institutional adoption of embodied GenAI tutoring systems, highlighting strategies for ethical deployment and inclusive design to maximize their socio-cultural benefits in education.
\end{itemize}

\section{Related Works}
\label{sec:related_works}

Recent advances in LLMs have catalyzed the development of intelligent 3D avatars capable of natural conversation in immersive environments~\citep{fink2024ai,wan2024building,zhao2024embodied,zhang2025effects,gao2025pervrml}. Early work by \citep{yamazaki2023open} integrated an LLM into an open-domain VR avatar chatbot, illustrating both the potential for free-form dialogue and the technical hurdles of achieving real-time multimodal synchronization (speech, gestures) without incurring high latency. Building on such foundations, \citep{wan2024building} created an LLM-driven agent within a social VR platform (VRChat) that combined GPT-4-based dialogue with long-term memory and retrieval modules to simulate human-like interactions. Their system addressed context management by optimizing the prompt length, striving to maintain conversational realism while controlling computational load. 

Another direction has focused on improving the interactivity and efficiency of these virtual agents through system design. For instance, \citep{maslych2024takeaways} deployed multiple voice-driven avatars in a VR scenario using a locally hosted LLM and an end-to-end interaction pipeline that included automatic speech recognition (ASR), text-to-speech (TTS), and real-time lip-syncing. 

These works have paved the way for LLM-driven Avatars adoption in education, where they serve as tutors, trainers, or guides. A recent work \citep{zhang2025effects} compares traditional text labels with LLM-empowered chatbots and avatar guides in a virtual museum to see how each affects visitor engagement, experience, and learning. Authors found that avatar guides powered by LLM significantly increase user engagement and improve the overall experience compared to both chatbots and static labels. However, all three interaction methods produce similar learning outcomes, showing that enhanced engagement and experience do not necessarily translate into greater measured learning.
In a pure educational setting, authors of \citep{neumann2024llm} performed a test on LLM-based chatbots deployed as virtual tutors in university courses. They showed how this supports personalized learning and self-regulated study, helping students ask questions and receive on-demand guidance inside learning management systems.
Finally, \citep{el2025architecture} introduces an architecture envisioning intelligent tutoring in immersive virtual reality that uses LLM-powered NPCs to interact with learners, combining adaptive dialogue with multimodal inputs like gesture and gaze. Experimental results indicate that such AI tutors enhance learner engagement and immersion compared to non-interactive environments. This work highlights the promise of LLMs for richer educational VR experiences while identifying technical challenges like real-time speech processing.

While the above works demonstrate the promise of LLM-driven virtual humans, many existing systems address only subsets of the overall challenge. In particular, prior solutions often trade off modularity or deployment flexibility for performance, or rely on cloud-based computation that introduces latency and privacy concerns. Our work aims at crossing these threads by providing a comprehensive framework that includes modularity, efficiency, and accessibility as main factors. For this, we designed a generalized framework for LLM-based avatars tailored to mainstream educational use and provided a first implementation of it. The proposed system is optimized (but scalable) for low-cost, on-device deployment, which not only yields low response times but also ensures that user interactions remain local to the device, providing privacy by design.

\section{ELEVATE: Efficient LLM Education with Virtual Avatar Teaching Engine}
\label{sec:elevateframework}

\begin{figure}[!h]
    \centering
    \includegraphics[width=1\linewidth]{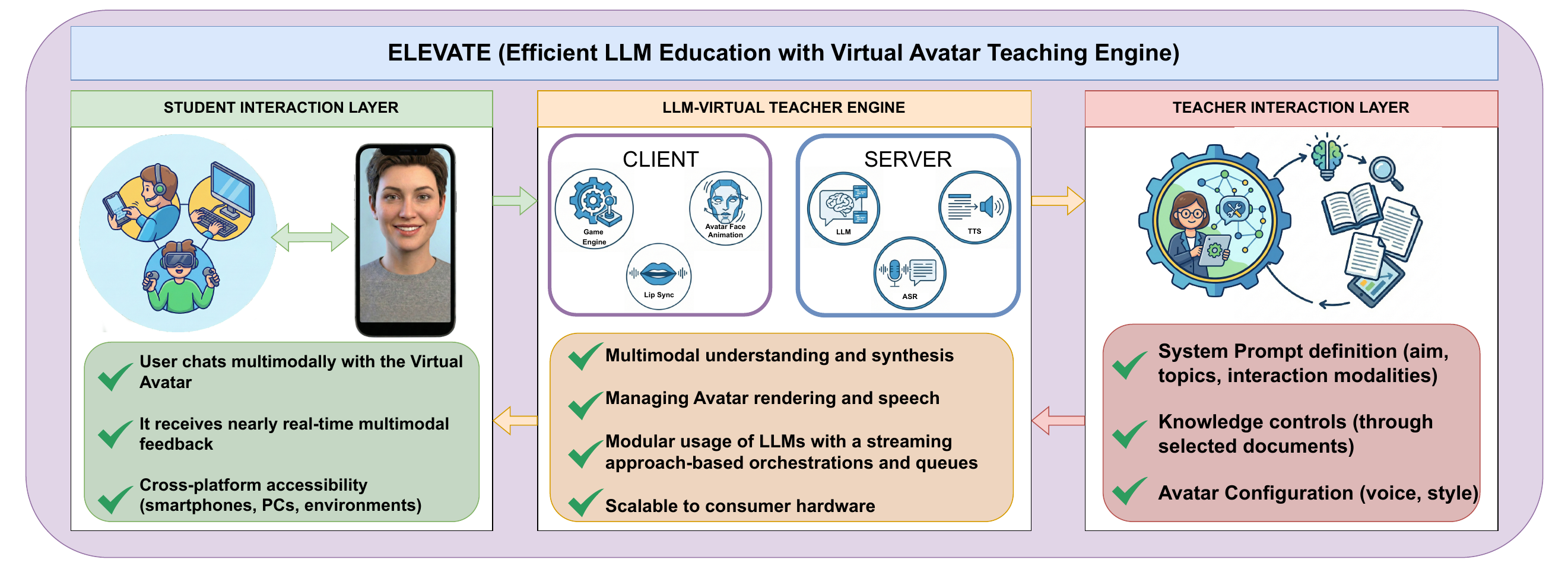}
    \caption{The ELEVATE framework overview is composed of three different strata: (i) Student Interaction Layer; (ii) LLM-based Virtual Engine; (iii) Teacher Interaction Layer.}
    \label{fig:elevateframework}
\end{figure}


The ELEVATE framework (Efficient LLM Education with Virtual Avatar Teaching Engine), visually depicted in Fig.~\ref{fig:elevateframework}, is conceptualized as a three-stratum loop designed to enable educational institutions to deploy GenAI–driven virtual tutors through an embodied, multimodal interface while preserving efficiency, privacy-by-design, and accessibility. The framework explicitly separates responsibilities across three tightly coupled but conceptually distinct strata: (i) a student-facing interaction layer, (ii) a local ELEVATE Engine core responsible for low-latency inference and multimodal synthesis, and (iii) a teacher-facing customization and governance layer.

This architectural design is motivated by the need to ensure reproducible classroom behavior, simplify deployment in real educational contexts, and clearly distinguish pedagogical intent (defined by educators) from system execution constraints (defined by hardware, models, and runtime parameters).
These three strata form a closed-loop system in which student interactions, generated through the student-facing layer, are processed by the ELEVATE Engine under teacher-defined constraints, and the resulting multimodal responses are embodied by the virtual avatar. This loop operationalizes a human-centered, efficient, and privacy-preserving approach to GenAI in education.
In the following sections, we detail the internal components and design choices of each stratum, highlighting how ELEVATE enables ethical, scalable, and pedagogically grounded deployment of LLM-driven virtual tutors in real-world educational settings.

\subsection{Student Interaction Layer} \label{subsec:stratum1_student} 

The first stratum of the framework (shown on the left side of Fig.~\ref{fig:elevateframework}) represents the learning interface. Students can interact multimodally with a virtual avatar using speech or text across multiple platforms, including smartphones, desktop PCs, and immersive 3D environments. This layer provides near real-time feedback through the avatar's synchronized speech, facial animation, and visual cues rendered on the device. The cross-platform design aims to make the same experience accessible regardless of device constraints, supporting inclusive and flexible learning scenarios. The main components of this strata amount to: 

\begin{itemize} 
\item \textbf{Cross-modal Avatar User Interface:} A unified interaction surface providing a dialogue panel, speech controls, and optional camera/microphone inputs (depending on device capabilities and school policies). The interface is designed to accommodate both typed and spoken interaction. 
\item \textbf{Avatar rendering:} A real-time 3D rendering subsystem built on established game-engine libraries, supporting avatar selection, animation control, lip-sync, gaze/gesture behaviors, and interaction-state cues (e.g., ``listening'', ``thinking'') that reduce perceived latency and clarify system state. 
\item \textbf{Accessible elements:} Core accessibility features include captions/subtitles, adjustable speaking rate, simplified-language toggles, and high-contrast UI variants. \end{itemize}

\subsection{ELEVATE Engine}
\label{subsec:stratum1}

The center stratum, the ELEVATE Engine, embodies the technical core of the framework and is logically divided into a client and a server component. The client manages avatar rendering, facial animation, lip-sync, and real-time interaction using a game-engine-based pipeline, while the server performs automatic speech recognition (ASR), LLM inference, and text-to-speech (TTS) synthesis. The engine adopts a modular and streaming-based orchestration strategy, enabling incremental generation and reducing perceived latency. 
It is worth noticing how such an engine must be local and scalable: it should be deployable locally or on-premise, scaling from consumer hardware without reliance on cloud-based services.

\subsubsection{Client-side responsibilities}
The client is designed to run on heterogeneous end-user devices (e.g., Android smartphones, desktop PCs, and XR-capable systems) and is responsible for (i) rendering the embodied interface, (ii) collecting multimodal input, and (iii) presenting multimodal output in a synchronized fashion. In particular, it orchestrates the 3D avatar runtime by composing standard game-engine modules (rendering loop, animation controller, blend shapes and bone-based facial rigs, audio playback) with conversational UI elements (text chat, microphone controls, ``thinking/listening'' indicators).
From a system perspective, the client provides three key functions: (1) interaction management, packaging user prompts and local context into structured messages; (2) embodiment and synchronization, mapping audio and control tags into lip-sync, gaze shifts, and turn-taking cues; and (3) accessibility rendering, ensuring that captions, reading-level adaptations, and alternative interaction modes remain available regardless of the back-end configuration.

\subsubsection{Server-side responsibilities}
The server represents the inference and synthesis back-end of the system and is intended to be executed either on a local machine within the institution or on an on-prem local network node. It is structured as a sequence of modular services that can be enabled, replaced, or scaled depending on the deployment profile and institutional constraints. Concretely, the server executes: (i) ASR to convert spoken prompts into text when voice interaction is enabled; (ii) LLM inference to generate pedagogically shaped responses; (iii) optional retrieval augmentation to ground responses in teacher-approved materials; and (iv) TTS synthesis to provide natural spoken output. The server also attaches lightweight metadata (e.g., segment IDs, timing information, and animation cues) to support synchronization on the client side.

\subsubsection{Engine Requirements}
A central design choice of ELEVATE is the adoption of streaming generation and incremental delivery. Instead of waiting for a full response completion, the server emits partial outputs (e.g., paragraphs or chunks) as soon as they are available. This reduces the perceived latency, supports conversational turn-taking, and aligns with embodied interaction requirements, where the avatar can display intermediate states (e.g., ``thinking'') and begin vocalization early. At the architectural level, this is achieved through queue-based orchestration between different modules (ASR, LLM, TTS), enabling asynchronous execution and preventing cascaded blocking that would otherwise increase end-to-end response time.

Moreover, ELEVATE is explicitly designed around a local-first principle: the inference pipeline can run without external cloud dependencies, which (i) reduces recurring costs, (ii) enables offline operation, and (iii) strengthens privacy-by-design by keeping student interactions within the institutional boundary. This is particularly relevant for European educational contexts, where data protection constraints and procurement limitations often make cloud-based LLM services difficult to adopt at scale~\citep{UNESCO2023GenAI}. In ELEVATE, institutional deployment can range from a single workstation serving a classroom to a local on-prem node supporting multiple devices, while preserving the same software abstraction and governance mechanisms.

Finally, the engine exposes explicit control surfaces to the Teacher Interaction Layer through parameters, allowing to control pedagogical intent as system specification that governs the LLM and retrieval behavior consistently across sessions and devices (as detailed in the following Section).


\subsection{Teacher Interaction Layer}
\label{subsec:teacher_layer}

The Teacher Interaction Layer (right side of Fig.~\ref{fig:elevateframework}) provides educators with an explicit control surface to configure and constrain the behavior of the virtual tutor. This stratum is conceived as a parameters editor, where system prompt, avatar configuration, and knowledge base are selectable and customizable (following guidelines from~\citep{knoth2024ai}).

\subsubsection{System prompt and pedagogical intent specification and Interaction policy}
A first set of controls allows teachers to define high-level instructional parameters that shape the conversational tutor at runtime. These include the educational aim (e.g., explanation vs.\ practice vs.\ formative assessment), target topic(s), grade level, tone and style (formal/informal), response length constraints, and interaction modalities (text-only, voice-enabled, or multimodal). In ELEVATE, these elements are compiled into structured prompt templates and policy rules consumed by the server, ensuring that pedagogical framing is persistent and not dependent on how a single student phrases a request.
Beyond content scope, the prompt allows explicit governance settings, including permitted task categories (e.g., explanation, quiz generation, feedback), refusal policies (e.g., disallowed requests), and safety filters appropriate for age and context. These controls are enforced server-side as runtime constraints, enabling teachers and institutions to define boundaries around academic integrity, sensitive topics, and acceptable assistance levels (e.g., hints vs.\ full solutions). This is especially relevant for assessment scenarios, where policy-driven behavior helps distinguish legitimate support from answer delivery.

\subsubsection{Avatar configuration and interaction style}
Teachers can configure the avatar persona to match classroom norms and learner needs. This includes voice selection and speaking style, as well as higher-level persona attributes (e.g., supportive vs.\ neutral stance, degree of scaffolding, and communication pacing). These parameters allow educators to set consistent interaction expectations and reduce variability across learners, which is particularly relevant in inclusive and heterogeneous classrooms.

\subsubsection{Knowledge governance: document selection, RAG, and document-driven adaptation}
A central feature of this layer is knowledge control through teacher-approved documents. Educators can curate a corpus of materials (e.g., textbook excerpts, lecture notes, worksheets, institutional policies, or open educational resources) that delimit the knowledge boundary of the system. These documents can be used by the server in two complementary ways:
(i) through Retrieval-Augmented Generation (RAG), where the server retrieves relevant passages from the curated corpus at query time to ground responses in approved sources; and (ii) through document-driven adaptation workflows, where the same materials can support lightweight fine-tuning or continual adaptation procedures (when feasible and permitted by institutional policy) to align the model with local curricula and terminology.

It is worth noticing that both mechanisms aim to improve factual consistency, reduce hallucinations, and align outputs with institutional expectations, while preserving auditability (e.g., tracking corpus versions and update workflows).
Finally, all teacher-defined settings could be stored as configuration artifacts (templates, rule sets, and corpus descriptors) that the server interprets at runtime. This makes the system behavior reproducible and transparent: changes in the tutor’s outputs can be traced to explicit modifications in prompt templates, policies, or document corpora, rather than to implicit prompt drift. In this sense, this stratum acts as the main interface through which educational institutions can exercise control over GenAI deployment, ensuring that adoption remains aligned with pedagogical goals, accessibility requirements, and ethical constraints (privacy-by-design, local governance, and accountability).

\section{ELEVATE Implementation}
\label{subsec:sysimpl}
In this section, we detail an instance implementation of the proposed ELEVATE framework.
We selected Unity game engine (v.6.0) and C\# with the .NET internal framework for the first stratum client building, a lightweight web-based application for the second-stratum configuration interface, and Python(v.3.13) for the third, computationally intensive stratum.
This toolchain was chosen primarily to maximize reproducibility across heterogeneous target devices, rather than to satisfy constraints intrinsic to the proposed design. On the client side, Unity provides mature cross-platform deployment support and enables high-quality real-time 3D rendering on diverse hardware, drawing on a long-standing engineering ecosystem for cross-device delivery. On the server side, Python was adopted to prioritize adaptability and maintainability in content generation and orchestration, as it supports rapid iteration on model integration, prompt and configuration logic, and incremental updates without tight coupling to a specific runtime stack.

This combination represents one instance of the possible languages and software stacks: the architecture was designed to remain adaptable across alternative implementations. In particular, the client runtime and the asset production pipeline can be realized through fully open-source solutions (e.g., Godot for real-time rendering and interaction) when institutional policies or educational-ethics requirements call for auditable and vendor-independent tooling. Conversely, the same module boundaries can also be integrated with paid, third-party APIs for specific capabilities; however, by design and to remain aligned with the authors’ ethical positioning, this is treated as an optional deployment path rather than as a requirement of the framework.

\subsection{Student Interaction Layer implementation}

\begin{figure}[!h]
    \centering
    \includegraphics[width=0.9\linewidth]{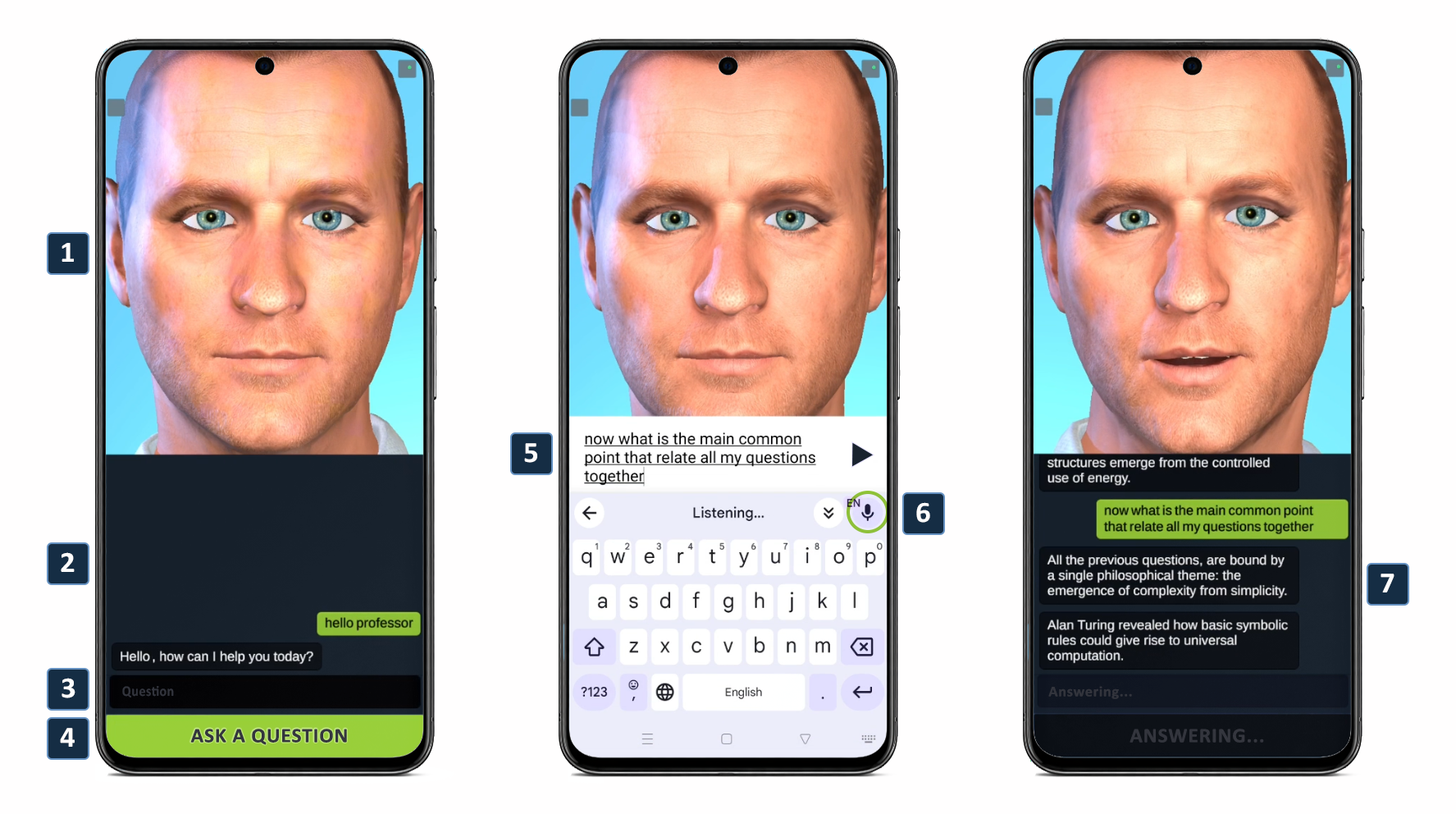}
    \caption{User interface components and interaction flow of the Student Interaction Layer, designed as a mobile application.}
    \label{fig:annotated-UI}
\end{figure}

The first stratum is implemented as a set of client-side components that jointly provide the learner-facing interaction surface for the target small-school scenario. Concretely, this stratum is responsible for (i) acquiring multimodal user inputs (typed text and, when permitted, speech), (ii) maintaining an explicit interaction state that is exposed to the learner through UI and avatar cues, and (iii) rendering the system’s responses through synchronized audio-visual behaviors (speech playback, facial animation, and interaction-state indicators). The implementation is organized to preserve a stable functional interface across heterogeneous execution targets (mobile devices, desktop PCs, and immersive 3D runtimes), while enabling platform-dependent capability gating driven by device constraints and institutional policies. At runtime, the first stratum operates as a thin, latency-sensitive client that orchestrates input capture, request/response presentation, and real-time avatar rendering, and that exposes accessibility controls as first-class configuration options. Fig.~\ref{fig:annotated-UI} provides a visual walkthrough of the client interface across the main interaction states.

Fig.~\ref{fig:annotated-UI} illustrates the learner-facing interaction loop as it unfolds across three interface states. In the initial state, the dialogue window (2) is present but not yet populated. Students can formulate a prompt either by typing in the input field (3), which triggers the device-native text entry widget (5), or by using the microphone control (6) when speech input is permitted. Once the prompt is submitted through the “Ask” action (4), the client presents the Virtual Teacher’s response through synchronized avatar delivery in the 3D view (1) and records the exchange in the dialogue window, which transitions to the scrollable history view (7).


\subsubsection{Interface Design}
The user interface constitutes the primary human-machine interaction surface of the system, as it mediates both prompt acquisition and response presentation. In our implementation, user prompts are acquired through two complementary modalities (typed text and speech input via device-native speech-to-text), which are implemented through a native UI layer (more details in Section~\ref{subsec:elevate_engine_imp}). Text input is captured through a native UI field integrated with the device input method, while speech input is exposed through a dedicated UI control that invokes the device-native speech-to-text subsystem and forwards the resulting transcript as the interaction prompt. This organization keeps the prompt representation uniform at the client boundary, since both modalities ultimately produce a textual query that can be dispatched to the downstream pipeline.

\par\medskip
System responses are presented through two core output channels: a 3D avatar visualization with lip-synched spoken delivery, and a dialogue window maintaining a turn-by-turn textual history. In our implementation, the virtual teacher is presented within a 3D scene, where speech playback is coupled with mouth movements (lip-sync) and can be paired, according to the educational needs, with complementary non-verbal behaviors defined in the Virtual Teacher personality, which are treated as first-class elements of educational communication. Voice timbre, speaking rate, and teacher personality traits can be surfaced to the student as part of response delivery, and can be parameterized through the system configuration layer described in the second stratum. In parallel, the dialogue view is realized as a native UI panel that maintains and renders the conversation history, and can additionally expose auxiliary elements required by the deployment context, including subtitles and other accessibility options, as well as operational controls such as audio/video recording icons and access points for downloading school-governance materials, when these features are enabled by configuration and policy.

The 3D avatar is designed from the outset for educational dialogue, so the client experience is intentionally framed as a video-call–like and face-centric Virtual Teacher, making facial articulation the primary communication channel and the surrounding 3D context secondary (as depicted in Fig.~\ref{fig:annotated-UI}). 
Finally, the client enforces clear turn-taking through listening/thinking/speaking states that mainly regulate interaction availability and convey state via audio-visual delivery and a persistent dialogue history, motivating the next section detailing the avatar implementation choices.

\subsubsection{3D Avatar Design and Asset Production Pipeline}
\label{subsec:avatar_design}

The Virtual Teacher avatar used in this prototype was manually authored by the authors through a dedicated asset-production pipeline, rather than being sourced from a pre-packaged character library. This choice was motivated by two complementary needs: (i) pedagogical control over the teacher’s appearance and expressive affordances (e.g., facial rig quality, clarity of lip movements, neutrality of visual style), and (ii) system integration requirements related to real-time rendering, animation retargeting, and low-latency embodiment on consumer hardware. In the context of ELEVATE, the avatar is not an ornamental element but part of the interaction channel: facial expressions, turn-taking cues, and lip synchronization are treated as communicative signals that shape how students interpret the tutor’s epistemic stance and instructional intent. The reference toolchain adopted in this work follows a conventional yet robust game-ready character pipeline:
\emph{Character Creator (v.4.4)} $\rightarrow$ \emph{ZBrush (v.2025.3.0)} $\rightarrow$ \emph{Substance Painter (v11.0)} $\rightarrow$ \emph{Mixamo humanoid auto-rigging and animation} $\rightarrow$ \emph{export (OBJ)}.
The pipeline can be decomposed into five stages:

\begin{enumerate}
    \item \textbf{Character Creator}: we first generated a neutral, classroom-appropriate teacher model with a controllable facial morphology and proportion constraints suitable for educational contexts. This stage establishes topology consistency and base mesh properties that later affect deformation quality (especially around the mouth and eyes).

    \item \textbf{High-frequency sculpt refinement:} the character was refined to improve facial landmarks and silhouette clarity. Although stylization could be used to reduce uncanny effects, we opted for a conservative design intended to remain legible on small screens (smartphones) while avoiding overly realistic features that may increase the risk of discomfort in some learners.

    \item \textbf{Material and texture authoring:} we authored physically-based materials (albedo, normal, roughness/metalness as applicable) to achieve coherent shading under diverse lighting conditions. Texture authoring was balanced against performance constraints: resolution was selected to preserve facial readability while limiting VRAM and bandwidth overhead for mobile/consumer devices.

    \item \textbf{Rigging and animation:} the character was automatically rigged to a standard humanoid skeleton to enable broad compatibility with common animation sets. Mixamo was used to obtain baseline idle and conversational animations and to simplify retargeting. In ELEVATE, this enables rapid iteration on embodied behaviors (idle stance, listening, speaking gestures) without requiring bespoke animation production at each iteration.

\end{enumerate}

It is worth noticing that, beyond visual elements, the avatar was designed under structural constraints. First, the face mesh supported clear lip articulation to avoid degrading comprehension during voice-based tutoring. At the same time, the asset must remain computationally tractable on consumer hardware: polygon count, texture sizes, and shader complexity were chosen to support stable frame rates under concurrency.

 \subsection{ELEVATE Engine Implementation}
\label{subsec:elevate_engine_imp}
\begin{figure}[H]
    \centering
    \includegraphics[width=\linewidth]{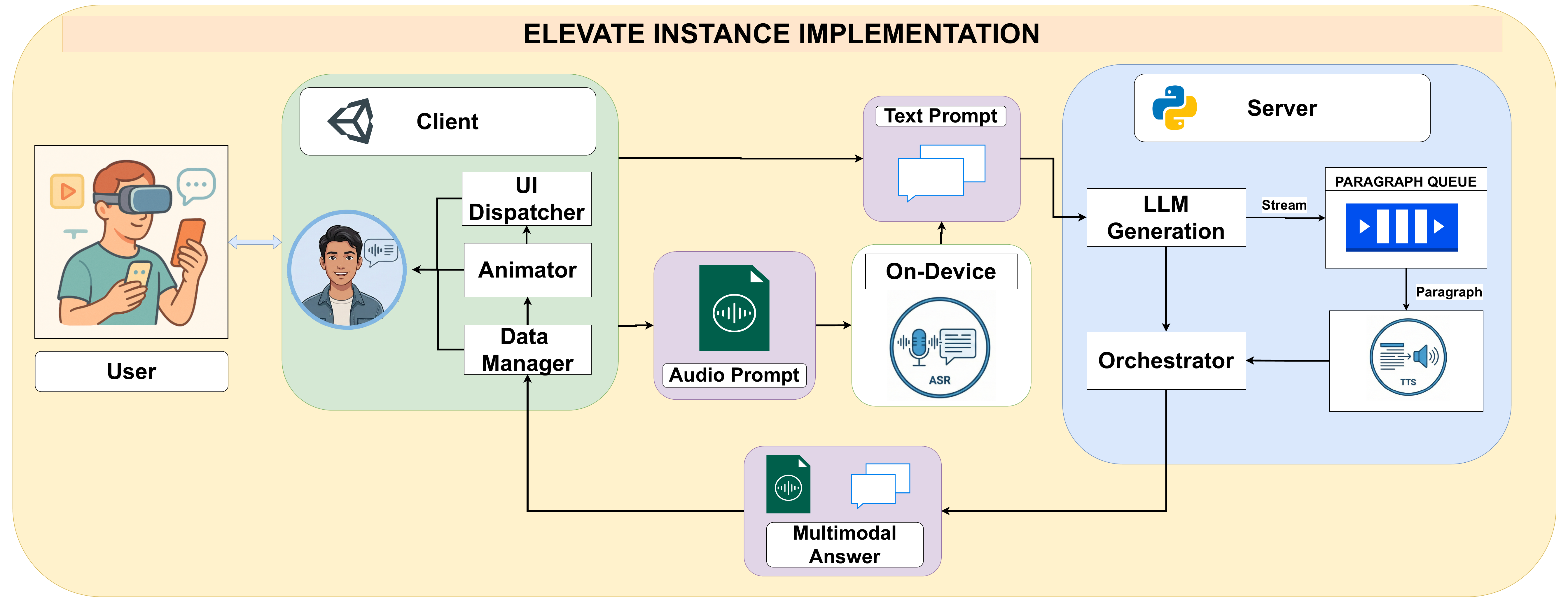}
    \caption{The ELEVATE implementation is a modular client–server pipeline in which the client manages user interaction, avatar animation, and multimodal input dispatching, while the server handles speech recognition, LLM-based reasoning, and text-to-speech synthesis, maintaining a scalable and loosely coupled orchestration of the AI components.}
    \label{fig:implementationengine}
\end{figure}

 The ELEVATE Engine represents the technical core of the framework and is implemented as a client–server pipeline that separates interaction rendering on heterogeneous end-user devices from generation and synthesis on local institutional hardware. Fig.~\ref{fig:implementationengine} shows the internal implementation of the engine. The client application is designed to run on resource-constrained platforms (e.g., smartphones, desktop PCs, and XR-capable systems) and is responsible for rendering the interface, managing interaction timing, and animating the Virtual Teacher. Conversely, the server hosts the computationally intensive modules, executing them on a local workstation or an on-premise node within the institution. This separation is adopted to keep the learner-facing runtime responsive while enabling the back-end pipeline to remain modular, maintainable, and deployable under local-first constraints.
 
In the reference implementation described in this section, the interaction boundary between client and server is text-based. 
The learner can provide prompts through multiple input options available on the client, including typed text, spoken interaction, and alternative or assistive input devices connected to the device, thereby supporting equitable access for students with diverse needs. When voice input is used, transcription is performed through device-native speech-to-text on the client side, and the server receives only a textual prompt.
Accordingly, the back-end pipeline does not include an explicit server-side speech recognition module in this implementation. The server instead focuses on secure ingestion of text prompts, incremental response generation, and local speech synthesis, returning a multimodal answer composed of both text and audio as a single encrypted payload.

\subsubsection{Client-side structure}
The client-side was implemented in the Unity Game Engine (in C\# with .NET support) and organized as a set of cooperating modules that jointly manage prompt acquisition, secure communication, and synchronized multimodal presentation (left side of Fig.~\ref{fig:implementationengine}). To keep the back-end interface stable across devices and permission regimes, all user inputs are normalized into a textual representation before leaving the device. This choice also allows the client to leverage OS-level accessibility affordances (e.g., native dictation, keyboard features, and assistive input methods) that are tightly integrated with mobile and desktop interaction.

The client runtime is structured into four main components:
\begin{itemize}
    \item \textbf{Native Speech-to-Text (STT):} When voice interaction is enabled, spoken prompts are converted into text using the device-native speech stack. We adopt native on-device STT to (i) minimize end-to-end latency and preserve interactive turn-taking, (ii) reduce server load and infrastructure cost by shifting speech processing to user hardware, and (iii) strengthen privacy-by-design by avoiding the transmission of raw audio to remote services.
    \item \textbf{Data Manager:} The Data Manager handles outbound and inbound messaging, packaging prompts into structured JSON requests and maintaining integrity, authentication, and ordering metadata. It also orchestrates the input/output flow between components, ensuring that partial responses arriving under streaming delivery are reassembled coherently and dispatched in the correct sequence.
    \item \textbf{UI Manager:} The UI Manager generates and updates the conversational interface at runtime. It collects user prompts (typed or transcribed), forwards them to the Data Manager, and renders incoming multimodal outputs (text plus playback state cues), maintaining a scrollable, turn-by-turn dialogue history.
    \item \textbf{Animator:} The Animator synchronizes avatar facial and body animation with received speech and text. In our reference implementation, the Virtual Teacher is imported into Unity as a rigged FBX model with blend shapes; we employ precomputed lighting to stabilize visual quality while reducing runtime rendering cost on consumer devices.
\end{itemize}

Outbound and inbound communication is managed by the Data Manager over a persistent Secure WebSocket (WSS) channel. In addition to transport-level security, the reference implementation applies authentication and application-layer encryption to exchanged payloads. On the outbound path, the Data Manager attaches required identifiers and integrity fields, then encrypts the JSON message prior to transmission. On the inbound path, it validates message structure and integrity, decrypts the payload, and only then dispatches the content to the UI and animation subsystems. Treating validation/decryption as an explicit stage of the interaction loop ensures that rendering logic operates exclusively on authenticated, well-formed content, and that malformed or unexpected messages cannot directly trigger presentation behaviors.

Concerning now the presentation layer, it couples a conversational UI with real-time avatar delivery. Upon reception of a response, the UI Manager updates the dialogue panel while the audio is played back, and the Animator aligns facial articulation and expression with speech. In the reference implementation, lip-sync is driven directly by the played audio, and the dialogue panel records each exchange as discrete turns to preserve instructional coherence. The client also enforces a strict turn-taking model: while the Virtual Teacher is speaking, prompt submission is disabled, and interaction resumes only after delivery completes. This constraint is deliberate: it simplifies synchronization across text rendering, audio playback, and avatar animation, and it reflects a “one speaker at a time” instructional exchange that improves interaction clarity for learners.

Because the server can deliver content in partial chunks, correct sequencing and synchronization are critical on the client side. The Data Manager is designed to accept streaming outputs and render them coherently as they arrive by relying on per-segment identifiers and ordering metadata. This ensures that dialogue updates, audio playback, and animation triggers remain aligned despite incremental delivery. The resulting synchronization logic (central to perceived latency and to the legibility of the embodied interaction) is described in detail in Section~\ref{subsec:queue_opt}.

\subsubsection{Server-side structure}

The server-side implementation is designed to execute \emph{within the institutional boundary}, either on a local workstation or on an on-premise node reachable through a local LAN/Wi-Fi. It is structured as a sequence of modular services, each implementing a single stage of the generation and synthesis pipeline (right side of Fig.~\ref{fig:implementationengine}). At a high level, the server receives a text prompt, authenticates and validates it, generates a response with an LLM, performs local speech synthesis, and returns a multimodal payload to the client for synchronized rendering.

The back-end is organized into the following logical components:
\begin{itemize}
    \item \textbf{AUTH Gateway:} Terminates the secure channel, authenticates clients, enforces authorization rules, and applies application-layer encryption and integrity checks to payloads.
    \item \textbf{Message Validator:} Verifies schema compliance (required fields, types, bounds), checks request identifiers and ordering metadata, and rejects malformed or unexpected messages before they enter the generation pipeline.
    \item \textbf{LLM Inference Module:} Executes local LLM inference and produces textual responses. This module is designed to support streaming generation and to attach per-segment metadata when incremental delivery is enabled.
    \item \textbf{Session Context Manager:} Maintains short-lived conversational state within a session scope (e.g., current turn, temporary context window), without persisting user identity or long-term transcripts by default.
    \item \textbf{Local Text-to-Speech (TTS):} Converts generated text into speech using a local synthesis engine (no cloud calls), returning audio plus optional timing/segment metadata.
    \item \textbf{Response Packager:} Constructs the JSON response payload, including text, audio, and ordering/synchronization metadata (message IDs, segment indices, timestamps) required by the client to align UI updates, playback, and animation.
\end{itemize}

At the ingress boundary, the AUTH Gateway manages authentication and application-layer encryption. Incoming client messages are admitted to the pipeline only after they are verified to originate from an authorized endpoint and validated against the expected schema. On the outbound path, the same gateway encrypts and signs the response payload prior to transmission. While the transport channel is secured, the implementation adds application-level protection. This also enables an explicit validation/decryption step on the client before any content can trigger rendering behavior.

Once a request is validated, the prompt is forwarded to the LLM Inference Module, which generates the textual response under the current session constraints. In the reference implementation, conversational context is maintained \emph{only within a session scope} and is not persisted across client restarts. Consequently, the server does not require persistent user identity management or cross-session conversation tracking. This boundary is intentional for the presented prototype: it keeps the system responsive and operational in classroom settings while avoiding assumptions about institutional identity infrastructures and long-term storage policies. The same modular boundary can later support policy-controlled persistence (e.g., teacher-approved logging or curriculum-bound memory) if required by specific deployments.

The generated text is then passed to the local TTS component, which produces speech without invoking cloud-based services. In ELEVATE, local synthesis is treated as a deployment constraint rather than a mere optimization: it reduces connectivity dependence, keeps audio within the institutional boundary, and provides predictable latency when running on provisioned local hardware. After synthesis, the Response Packager constructs a structured JSON object containing (i) the response text, (ii) the corresponding audio payload, and (iii) lightweight metadata required for ordered delivery and synchronized playback (e.g., message identifiers, segment indices, and timestamps). When streaming delivery is enabled, the same mechanism supports partial responses by emitting chunked segments that the client can reassemble coherently.

To conclude, the pipeline is designed so that individual stages can be enabled, replaced, or scaled depending on the deployment profile. In the small-school reference implementation, the back-end focuses on secure prompt ingestion, local LLM generation, local speech synthesis, and encrypted multimodal delivery. Additional capabilities (such as RAG over teacher-curated documents, teacher-facing policy enforcement or domain adaptation via controlled fine-tuning) can be integrated by attaching modules at the validated request boundary and the session/context layer, without altering the core client-server interface. This preserves interoperability across devices while enabling institutions to progressively increase capability under their own infrastructural and governance constraints.

\subsubsection{Asynchronous data management and streaming}
\label{subsec:queue_opt}

A central requirement of the ELEVATE Engine is support for low perceived latency through asynchronous orchestration and streaming delivery. Rather than waiting for full response completion, the server emits partial outputs as soon as they become available (e.g., paragraph-level or chunk-level segments). This enables the client to populate the dialogue panel incrementally and to begin response delivery earlier. From an interaction standpoint, streaming supports smoother turn-taking, as the learner receives timely feedback and the system can proceed with delivery without a long silent interval.

 Architecturally, streaming is realized by decoupling pipeline stages through queue-based orchestration (as also depicted in the right part of Fig.~\ref{fig:implementationengine}). The LLM component produces text in segments that are placed into intermediate queues, where downstream components, TTS synthesis and delivery, consume them asynchronously. This arrangement avoids cascaded blocking across the pipeline: generation can progress independently of synthesis, and packaging/delivery can transmit segments as soon as they are completed. The response is then delivered either as an ordered sequence of encrypted payloads or as a single encrypted payload containing ordered segments, depending on the deployment configuration, together with metadata that allows the client to reassemble and present the output consistently.
This design provides an additional systems-level advantage: the server pipeline is \emph{deployment-agnostic} and can be \emph{horizontally partitioned} across multiple nodes. In particular, the Orchestrator, LLM inference, and TTS services can be executed as separate processes on different machines, communicating through the same message interface. This distributed deployment model was introduced to reflect realistic school infrastructures, where resources are often available as multiple low-end consumer devices rather than a single high-performance workstation. By allowing compute-intensive stages to be offloaded to different nodes, the architecture improves scalability and preserves responsiveness without requiring centralized, GPU-grade hardware.

 However, such an asynchronous organization requires explicit coordination mechanisms on both client and server. In the reference implementation, this coordination is provided by internal asynchronous modules, each responsible for a narrowly scoped function and communicating through lightweight queues and shared state. On the client side, these modules cover message transmission and reception, ordering, validation/decryption, UI updates, and audiovisual synchronization; on the server side, analogous modules handle request validation, generation, synthesis, packaging, and delivery. This decomposition is essential in asynchronous settings because queue management (including ordering, partial delivery, and failure isolation) cannot be reliably sustained by a single monolithic control loop. The system was conceived by the author to mirror procedures typical of bureaucratic environments, where both the intake clerk and the issuing clerk must validate documents through matching stamps, and only when these stamps correspond is a document treated as valid; analogously, requests and responses are admitted to downstream stages only after their validation and integrity checks succeed. By coordinating responsibilities through explicit queues, the engine remains controllable under streaming conditions and can evolve without entangling interaction logic with transport and scheduling details.

 Finally, the streaming design is paired with explicit security handling. Each partial output is subject to the same authentication, validation, and encryption constraints as a full response, and the client validates and decrypts segments before rendering. This ensures that incremental delivery does not weaken the integrity of the interaction loop and that the learner-facing client remains robust even when responses are delivered as multiple ordered units.

\subsection{Teacher Interaction Layer}

\begin{figure}[H]
    \centering
    \includegraphics[width=0.9\linewidth]{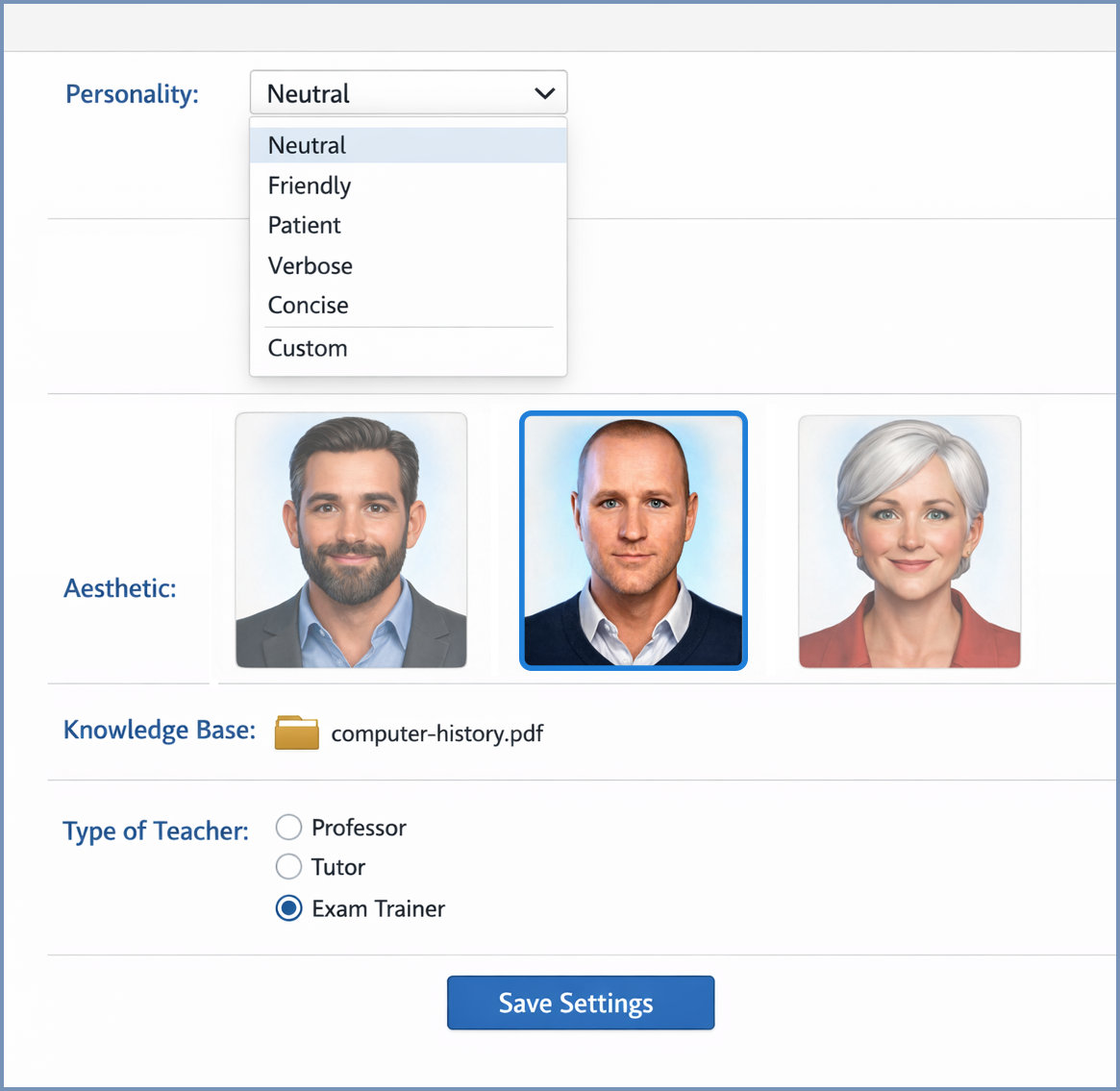}
    \caption{Teacher Interaction Manager}
    \label{fig:teacherinteractionmanager}
\end{figure}

The Teacher Interaction Layer is conceived as a lightweight, web-based configuration page through which educators (or institutional governance) can select a small set of high-impact parameters that shape the Virtual Teacher’s behavior (it is worth highlighting that all this information is injected directly into the LLM system prompt).
Fig.~\ref{fig:teacherinteractionmanager} illustrates an example design layout, organized as a single configuration page intended to provide an explicit control surface for pedagogical framing, interaction style, and knowledge boundaries.

In our design, the first control group concerns personality, implemented as a drop-down menu that exposes a deliberately constrained set of presets (e.g., \textit{Neutral}, \textit{Friendly}, \textit{Patient}, \textit{Verbose}, \textit{Concise}), so that educators can select an interaction style without having to specify low-level behavioral rules. The menu also includes a “Custom” option, which makes the configuration surface extensible beyond the predefined presets. In practice, this enables maximal flexibility in adapting the Virtual Teacher’s behavior to specific institutional or classroom needs: when a desired interaction style cannot be expressed through simple parameter selection, it can be realized through additional engineering inputs (e.g., tailored prompt templates, policy rules, or dedicated behavioral constraints) while preserving the same overall configuration workflow.

A second group addresses the avatar’s aesthetic configuration, enabling teachers to select the Virtual Teacher’s appearance through visual thumbnails rather than through textual labels. In the illustrated interface, candidate avatars are shown as portrait cards, and the active selection is indicated through an explicit visual state (e.g., a highlighted border). This presentation keeps the selection process concrete and low-effort, while avoiding reliance on potentially value-laden descriptors: by centering the choice on images instead of categorical wording, the interface supports a more diplomatically neutral and gender-respectful configuration workflow. Operationally, the thumbnail-based design also reduces ambiguity and setup time when configuring multiple classroom profiles, since educators can compare alternatives quickly and apply consistent selections with minimal interaction overhead.

The configuration page further includes a knowledge base selector that lists teacher-approved documents (e.g., a relevant pdf, like \textit{computer-history.pdf}) as the curated material the system may draw upon in deployments that adopt document-governed generation. This section is also the natural entry point for uploading and maintaining the reference texts that the institution intends the Virtual Teacher to rely on, so that explanations can be aligned with the exact exposition adopted in the classroom (e.g., the same terminology, definitions, and sequencing used in lectures or textbooks). In this sense, the knowledge base acts as an explicit boundary-setting mechanism: it makes the set of admissible sources visible at configuration time and supports subsequent runtime grounding in approved materials, thereby reducing ambiguity about “what the system is allowed to know” within a given educational context.

Finally, the interface specifies the type of teacher through mutually exclusive radio-button options (e.g., \textit{professor} (replicating lecture content), \textit{tutor} (providing repeated explanations and scaffolding), and \textit{exam trainer}), prioritizing questions and quizzes for assessment-oriented practice. The configuration flow is closed by an explicit save settings action, which indicates that the selected parameters are intended to be persisted as a coherent profile rather than applied transiently.
Selected settings can be stored as configuration artifacts (e.g., templates, rule sets, and corpus descriptors) interpreted by the server at runtime, enabling traceability of behavior changes to explicit configuration edits.

%

\section{Experiments and Results}

We here report the performed experimental evaluation to demonstrate the practical deployability and near real-time performance of the ELEVATE implementation under realistic school constraints (employing multiple consumer-grade hardware commonly available in public institutions, including standard PCs and smartphones).
The experiments focus on system-level performance indicators that directly affect classroom usability: end-to-end latency, time-to-first-talk and perceived latency. We also report the hardware setting adopted to run the entire system. Finally, we provide a technical discussion on limitations and scalability across heterogeneous deployment profiles.

\subsection{Apparatus and Configuration}
\label{subsec:app_conf}
For the implementation of our system for the small school scenario, we adopted a minimal hardware configuration required to achieve full real-time performance. 

\paragraph{Hardware Configuration}
Considering a real-world setting (i.e., schools with many consumer hardware), we exploited the ``partioning'' property of our server implementation (as detailed in Section \ref{subsec:queue_opt}), and we executed our system (and related experiments) in two local machines: (i) used to run the LLM generation and Orchestrator modules, was equipped with an AMD Ryzen 9 9900x CPU, 64 GB DDR4 RAM, and an NVIDIA RTX 5060 Ti with 16 GB of VRAM; (ii) where the TTS is executed, equipped with an Intel Core i7 6700K CPU, 32 GB DDR4 RAM, and an NVIDIA GeForce RTX 3060 with 12 GB of VRAM.
This configuration is capable of keeping both the LLM and TTS models fully resident in memory and performing low-latency inference (while adopting a low-cost and consumer device).
Given the extremely modular capabilities of ELEVATE, the reported results indicate that GPU technologies from different generations can coexist within the same deployment without introducing integration bottlenecks, which supports economically efficient rollouts and long-term maintenance by enabling incremental hardware reuse rather than synchronized full-system replacement. 
This approach allowed the entire server-side pipeline to run locally, without reliance on cloud infrastructure.
The Unity-based client application runs separately on target devices (a consumer Nokia G42 Android smartphone), which maintains a lightweight runtime profile thanks to the complete offloading of computational tasks to the server.

\paragraph{LLM and TTS Configuration}
As LLM, we selected Hermes-3B model~\citep{teknium2024hermes3technicalreport} running locally via llama.cpp~\footnote{\url{http://github.com/ggml-org/llama.cpp}}, and their respective audio renderings were synthesized through the CoquiTTS-based engine ~\footnote{\url{https://github.com/coqui-ai/TTS}} (same configuration adopted in our implementation).

We selected Hermes 3B after a hardware-constrained model screening aimed at deployment on consumer GPUs (with 8–16 GB of VRAM). Beyond this constraint, we evaluated a set of widely used small-to-mid LLMs that could plausibly fit within this memory envelope under quantization, and we probed each candidate using the same fixed question set to ensure comparability under identical prompting conditions. Under a controlled persona configuration, Hermes 3B consistently produced well-structured answers, with respect to parameter-comparable ones (i.e., llama3.2 3B). In contrast, larger models increased operational fragility by triggering VRAM saturation and unstable runtime behavior on the target class of devices. 
Following an Occam’s razor rationale, we therefore adopted Hermes-3B.Q8 as the default inference backbone \footnote{Q8 refers to the quantized 8-bit version}.
Finally, we employed such models native to llama.cpp with a custom llama.cpp build targeting the GPU CUDA stack Toolkit 12.9 to maximize end-to-end throughput and reduce perceived latency via efficient token streaming during sentence generation.

Coqui TTS was selected for two main reasons: firstly, it aligns with ELEVATE’s local-first, modular engineering goals while remaining easy to integrate on commodity hardware; secondly, it provides multiple integrations for programmatic orchestration, and a ready-to-deploy pretrained model, allowing users to choose smaller or more efficient voices to cope with school-grade hardware constraints.
Coqui TTS was deployed on GPU acceleration with CUDA runtime version 13.1, ensuring compatibility with the local GPU execution path while maintaining a reproducible, vendor-contained deployment profile suitable for institutional environments. Note that versions for all libraries, together with inference parameters, are detailed in the Supplementary File.

\subsection{Metrics}
\label{subsec:metrics}
To validate the runtime performance of our implementation, we measured inference latency across the server-side generation pipeline. Client-side latency was not explicitly measured, as it was negligible in practice: all client components are limited to orchestration, rendering, and playback operations, and do not perform computationally intensive inference or transformation steps.

In particular, we quantified the time required for LLM-driven sentence generation (split with end-of-sentence textual token) and the subsequent TTS synthesis (in seconds, named as $LLM_T$ and $TTS_T$). 
We also collected the number of tokens ($TK$) generated by the LLM, along with the concrete answers (which were all included in the Supplementary Material).
We report these measurements at this granularity to reflect ELEVATE’s incremental, streaming interaction design.

This also allows us to measure the Perceived Latency ($PL$) from the user, which depends on the cumulative delay introduced by each step of the pipeline (text generation, segmentation into sentences, TTS synthesis, and progressive delivery). We define the PL as:
\begin{equation}
PL_i = \max\!\bigl(0,\; (LLM_{T,i} + TTS_{T,i}) - S_{i-1}\bigr),
\qquad i = 1,\dots,N,\;\; S_0 \equiv 0.
\label{eq:perceived_latency_compact}
\end{equation}

In the equation we show the masking effect induced by incremental delivery: if the avatar is still speaking ($S_{i-1}$) while the next unit is being generated ($(LLM_{T,i} + TTS_{T,i})$), the user perceives zero or negligible waiting; conversely, if the generation exceeds $S_{i-1}$, the uncovered portion becomes an audible pause between spoken units. In our analysis, $PL_i$ is computed on mean timings per unit, consistent with the implementation used to derive the plotted perceived-latency overlays.

Moreover, we were able to calculate end-to-end latency and time-to-first-talk.
Let a response be streamed as $N$ spoken units (e.g., sentences/paragraphs) indexed by $i=1,\dots,N$.
For each unit, we measure the server-side generation time as the sum of LLM decoding and TTS synthesis,
$G_i = LLM_{T,i} + TTS_{T,i}$, and we denote by $S_i$ the playback duration of the synthesized audio for unit $i$ (with $S_0 \equiv 0$).
We define the \emph{end-to-end server latency} of a full conversational response as:
\begin{equation}
L_{\mathrm{E2E}} \;=\; \sum_{i=1}^{N}\bigl(LLM_{T,i} + TTS_{T,i}\bigr)
\;=\; \sum_{i=1}^{N} G_i .
\label{eq:e2e_latency}
\end{equation}

We define the \emph{time-to-first-talk} (TTFT), i.e., the time until the first audio segment is available for playback, as:
\begin{equation}
TTFT \;=\; LLM_{T,1} + TTS_{T,1} \;=\; G_1 .
\label{eq:ttft}
\end{equation}

\subsection{Real-World Use Case: History of Computer Science Tutor}

To demonstrate ELEVATE in a concrete yet broadly replicable scenario, we implemented a virtual teacher designed to support classroom dialogue on the \emph{History of Computer Science}. The use case targets short, curiosity-driven interactions (e.g., definitions, historical context, and conceptual clarifications). This domain was selected because it benefits from conversational explanation while requiring strong factual discipline: historical claims are easily verifiable.

\paragraph{System Prompt Configuration}
The system prompt is treated as a teacher-governed configuration artifact that operationalizes pedagogical intent at runtime. In this reference implementation, the prompt defines the avatar’s instructional persona, interaction format, and strict content constraints to minimize hallucinations and off-topic drift. The prompt used is reported below.
We intentionally adopted a simple,short and explicit prompt rather than a long, narrative persona description. This reflects practical constraints of local-first deployment with compact models. At the same time, it mimics what a non prompt-engineering expert (i.e., a Teacher) would do to set the behaviour of the avatar.
Constraining the generation of ``at least five sentences'' was selected for analysis reasons (see Section \ref{subsec:exp_prompt}).

\begin{figure}[!htbp]
\centering
\begin{tcolorbox}[
  colback=white,
  colframe=black,
  title=ELEVATE System Prompt,
  fonttitle=\bfseries,
  enhanced,
  attach boxed title to top center={yshift=-3mm},
  boxed title style={colback=black!80!white, colframe=black, size=small},
  boxrule=0.9pt,
  arc=1mm,
  top=4mm, bottom=4mm, left=4mm, right=4mm
]
\small
\begin{itemize}
    \item []
    \textit{Respond in English using a formal, academic professor style. The response must simulate a speech of a professor to a single student, and must consist of at least five sentences. The first sentence must be a single, short sentence of no more than ten words, with a neutral tone. Content constraints: the answer must be formal and academic, without labels and listings. Provide only correct, verifiable information strictly related to the history of computing. Do NOT invent facts, names, dates, or events. Avoid speculation, fictional content, or extrapolation beyond established historical sources. Never label each sentence. Failure to comply with any structural or content rule is a severe error.}    
\end{itemize}
\end{tcolorbox}
\end{figure}

\paragraph{LLM Model Adaptation to Domain Knowledge}
To keep the baseline architecture lightweight and reproducible in school settings, we intentionally adopted a RAG-free configuration and instead performed a one-time domain adaptation of a small local LLM. Specifically, we fine-tuned a 3B-parameter Hermes-3B model with a compact QLoRA procedure, then merged the learned adapters into the base checkpoint and converted the resulting model into an efficient local deployment suitable for offline inference (full training details are provided in the Supplementary Material). This training step was motivated by the need to inject domain knowledge into the tutor: by adapting the model directly on the target educational material, the system can answer topic-specific questions with less reliance on external retrieval services and with more consistent terminology and coverage for the intended curriculum. In practical terms, this approach reduces runtime complexity (no embedding/index maintenance) and shifts the computational cost to a one-time training phase, after which the model can be deployed locally without additional infrastructure.


\subsection{Experimental Setting and Results}
\label{subsec:exp_prompt}
To measure the selected metrics, we used a set of 20 representative student-style prompts (included in the supplementary material) to stress the end-to-end pipeline under realistic classroom interaction patterns, while keeping the experiment statistically valid, reproducible, and interpretable. As mentioned, we measured latency at sentence granularity (LLM generation + TTS synthesis) as soon as each sentence was ready, so prompt choice must reliably trigger multi-sentence explanatory outputs and heterogeneous generation paths. As mentioned, we constrained our LLM model to generate 5 sentences. 

\begin{figure}[!h]
    \centering
     \begin{subfigure}[t]{0.48\linewidth}
        \centering
        \includegraphics[width=\linewidth]{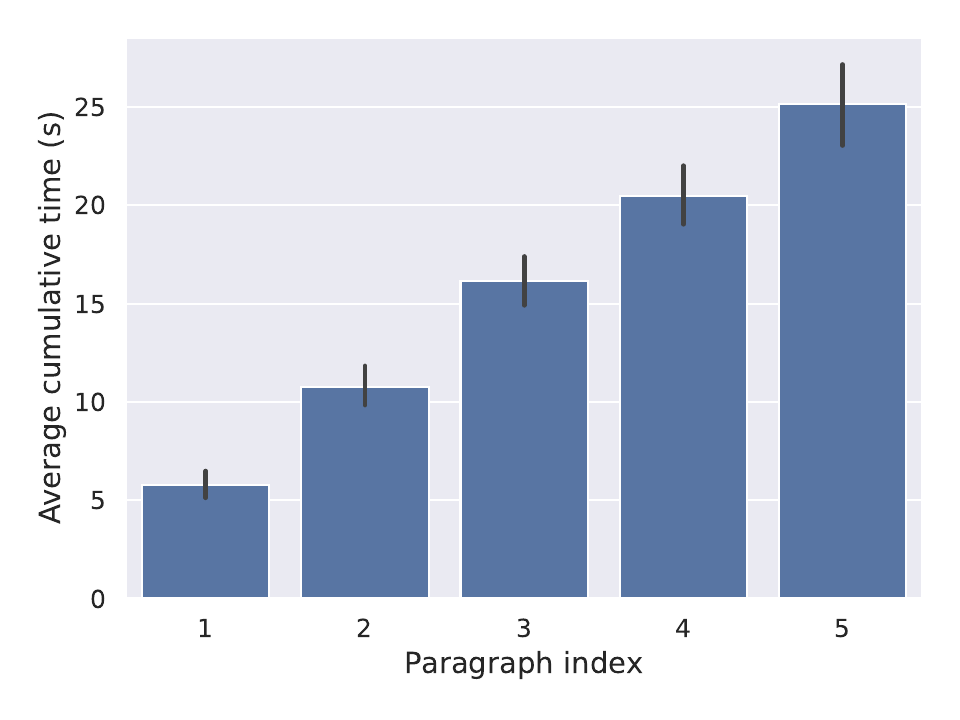}
        \caption{}
        \label{fig:cum_time}
    \end{subfigure}
    \hfill
    \begin{subfigure}[t]{0.48\linewidth}
        \centering
        \includegraphics[width=\linewidth]{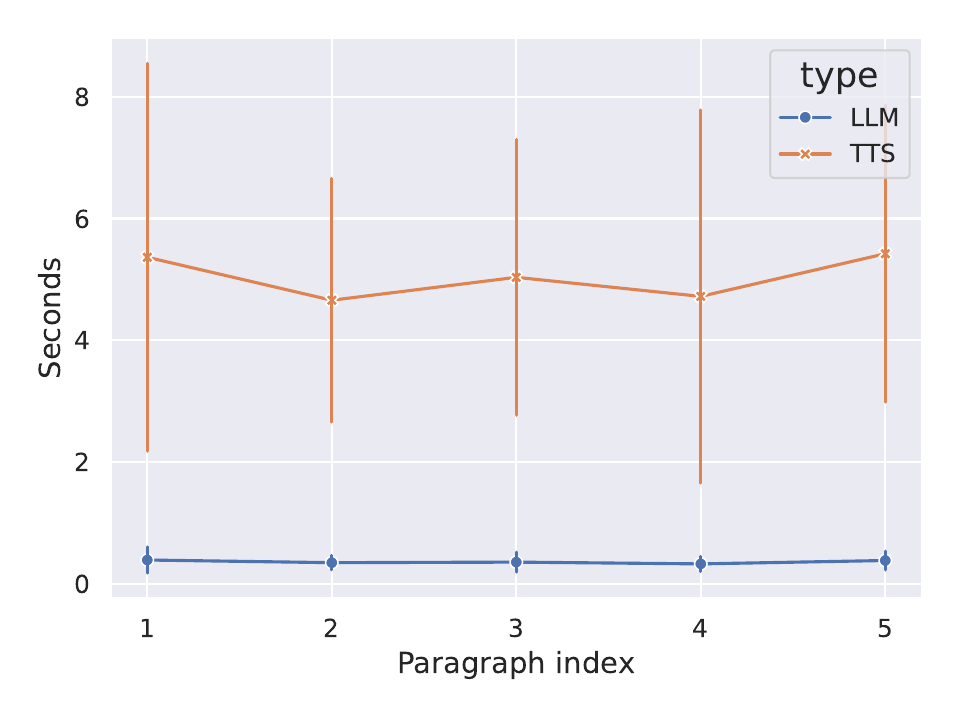}
        \caption{}
        \label{fig:avg_time}
    \end{subfigure}   
    \caption{Performance of the ELEVATE generation pipeline: (a) Average End-to-End Latency time as sentences are produced; (b) Average Per-Paragraph Generation Time, decomposed for LLM inference and TTS synthesis.}
    \label{fig:runtime_results}
\end{figure}

The results reported in Fig.~\ref{fig:runtime_results} indicate that the overall generation times are well within acceptable bounds for interactive educational use. In particular, the first bar in the cumulative plot (Fig.~\ref{fig:cum_time}) corresponds to the \emph{time-to-first-audio}, which represents the initial wait experienced by the learner before the virtual teacher begins speaking. This delay is approximately five seconds on average, which remains compatible with conversational and instructional interaction patterns in classroom and individual study settings. The error bars denote standard deviation across prompts, which shows a modest variability. The decomposition shown in Fig.~\ref{fig:avg_time} further highlights that LLM inference contributes only a minor fraction of the total latency. This confirms that the selected 3B-parameter LLM, executed on a more powerful GPU (RTX 5060), achieves fast and stable inference times even under streaming generation. By contrast, the dominant component of the pipeline latency is TTS synthesis. This is an expected outcome given that it is executed on a less powerful GPU and the selected library has no underlying optimizations (as detailed in Section~\ref{subsec:app_conf}) and reflects a deliberate deployment choice prioritizing modularity and resource reuse over peak audio synthesis performance. However, the standard deviation is high, which can be primarily attributable to differences in utterance length and phonetic complexity.

\begin{figure}[!h]
        \centering
        \includegraphics[width=0.7\linewidth]{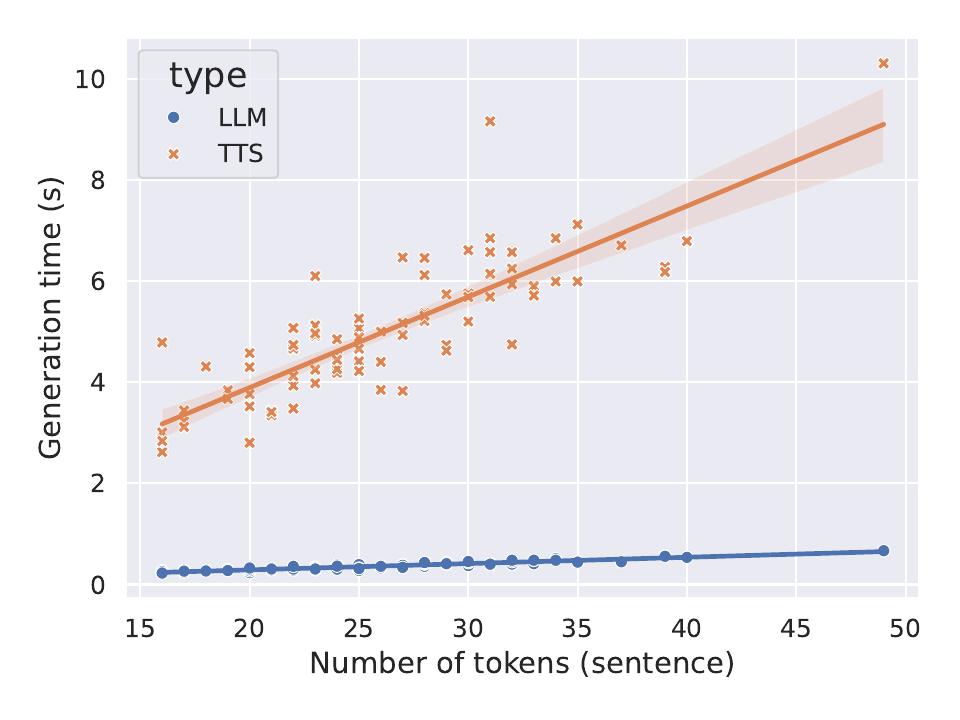}
        \caption{LLM and TTS generation time as a function of sentence length (number of tokens).}
        \label{fig:scatter_tokens}
\end{figure}

To further analyze the observed variability in synthesis time, Fig.~\ref{fig:scatter_tokens} reports a scatterplot relating sentence length (number of tokens) to generation time for both LLM inference and TTS synthesis.
The results confirm that LLM inference exhibits a weak and nearly linear dependence on token count, with consistently low absolute latency, highlighting the efficiency and stability of the selected 3B-parameter model under streaming conditions.
By contrast, TTS synthesis time shows both a steeper growth trend and a markedly higher dispersion. This behavior reflects the combined effect of increased utterance duration, phonetic complexity, and non-uniform workload on the shared, lower-tier GPU used for speech synthesis.

\begin{figure}[!h]
    \centering
    \includegraphics[width=0.8\linewidth]{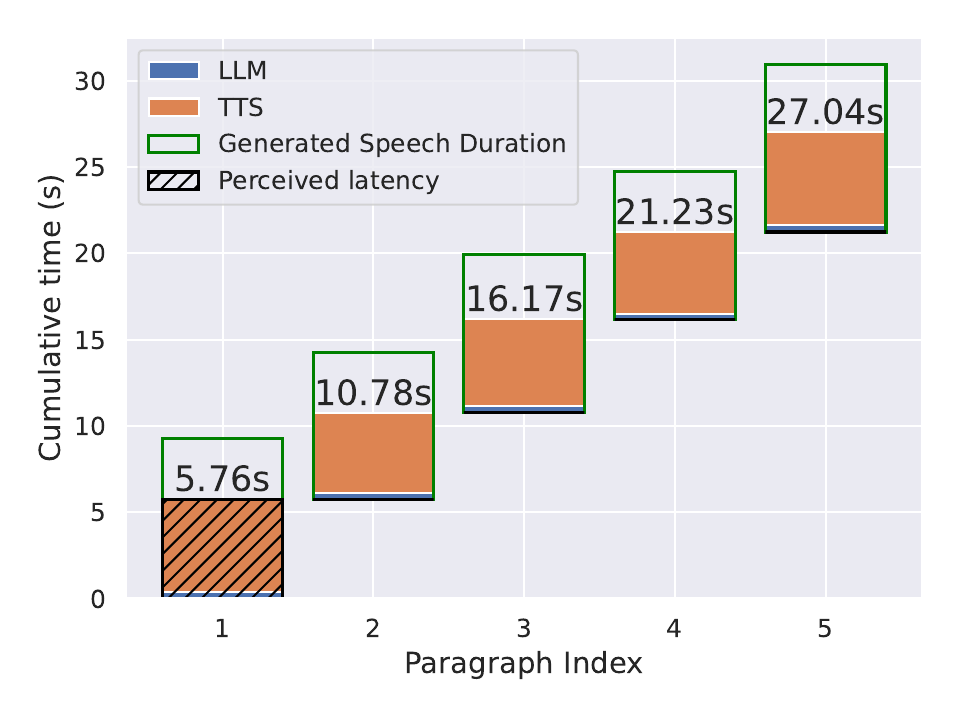}
    \caption{Cumulative decomposition of generation time across sequential paragraphs, showing the contributions of LLM, TTS, and speech duration. The hatched area highlights perceived latency, which is concentrated in the first paragraph (time-to-first-audio) and effectively masked in subsequent steps by streaming speech playback.}
    \label{fig:perceived_latency}
\end{figure}

However, from a user-experience perspective, the impact of this variance is mitigated by our streaming design of the pipeline, which allows speech playback to begin as soon as audio segments are available, decoupling perceived responsiveness from worst-case synthesis time. To demonstrate this effect, we analyze the perceived latency as defined in Section~\ref{subsec:metrics} and illustrated in Fig.~\ref{fig:perceived_latency}.

This plot decomposes cumulative generation time into its constituent components (LLM inference, TTS synthesis, and generated speech duration). However, it also explicitly highlights the portion of delay that is actually experienced by the user. As shown, the perceived latency is concentrated almost entirely in the first generation step, corresponding to the time-to-first-audio (approximately 5–6 seconds on average). After speech playback begins, subsequent generation and synthesis steps are largely masked by the ongoing audio output, resulting in a perceived latency that effectively drops to zero for later paragraphs.

\subsection{Technical Discussion and Limitations}

From a technical standpoint, the presented results support the feasibility of running an embodied, streaming GenAI tutor on consumer-grade devices; however, they also outline where the current baseline configuration reaches its limits. The empirical decomposition of runtime costs indicates that the pipeline is not uniformly bounded by LLM inference; instead, end-to-end responsiveness is dominated by speech synthesis and, more generally, by the heaviest stage executed on the least capable compute resource. This observation is consistent with the design choice of prioritizing modularity and local deployability over centralized, high-end acceleration, but it also surfaces concrete constraints that must be considered in real deployments.
In particular, under the baseline configuration, computational constraints may introduce delays when handling longer explanations. 
However, such limitations may be mitigated predictably with increased hardware capacity (key factor in ELEVATE). Provisioning a more capable on-prem server (or dedicating accelerators to TTS) directly reduces time-to-first-audio and stabilizes per-segment synthesis, enabling longer and more complex explanations without disrupting conversational flow. 

Another limitation involves the in-memory session of the conversation: lacking continuity across classes and chats may produce inconsistent behavior after client restarts; at the same time, adding persistence introduces governance requirements (retention, consent, access control).
Moreover, even with multiple local servers, the queue system may become a bottleneck when serving multiple students concurrently. Without a queueing policy and admission control limits, the actual system may be subjected to phenomenon like the ``noisy neighbor'' effects.

\section{Sociotechnical and Pedagogical Implications}
\label{sec:implications}

ELEVATE was introduced to address practical barriers that often prevent GenAI tutors from being adopted responsibly in schools: high operational cost, fragile connectivity requirements, limited institutional control over data and policies, and interaction modalities that do not map well to everyday classroom routines. By design, our platform targets \emph{deployability under real constraints} (consumer hardware, local execution, modular components) while preserving educator-centered governance through an explicit separation between student interaction, teacher control, and execution infrastructure. These design choices respond to the core challenges raised in this work: the need to integrate generative tutoring into education without outsourcing pedagogical authority, privacy obligations, and integrity policies to opaque third-party systems.

However, meeting efficiency and deployment constraints is only the entry point to responsible GenAI adoption in education: the most consequential challenges emerge once the system becomes part of the classroom ecology. An embodied, voice-based tutor can reshape participation norms, alter help-seeking patterns, and change how learners attribute authority and expertise; in parallel, local-first operation shifts control of data, models, and policies from external providers to schools, thereby relocating responsibility for safety, maintenance, auditing, and staff training onto the institution. Even with teacher-in-the-loop governance, key risks remain structurally open: hallucinations and epistemic unreliability, over-reliance and deskilling, uneven benefits across student groups, and the possibility that accountability mechanisms drift into surveillance through excessive logging. 
For these reasons, we treat ELEVATE not only as a deployable architecture but as a sociotechnical intervention whose educational value and ethical acceptability depend on how different stakeholders interact with it in practice. Accordingly, the remainder of this section is organized around the main actors and institutional functions that mediate these effects: we first analyze implications for students (experience, agency, educational identity) and teachers (role, pedagogy, and classroom orchestration), and then derive requirements for authorship and integrity, system assessment and control, and institutional adoption, governance, and inclusion.


\subsection{Implications for School Actors}
\subsubsection{Student experience, agency, and educational identity}
\label{subsec:student_agency}

From the learner’s perspective, an embodied GenAI tutor is not merely a utility but a social and epistemic actor. Always-available tutoring can reduce productive struggle, shift help-seeking away from teachers and peers, and change students’ sense of ownership over learning outcomes~\citep{constantinescu2025genai}. These effects may be amplified by embodiment through anthropomorphism and social bonding, especially for younger learners or those who prefer private, low-stakes interaction. As a result, responsible deployment requires explicit strategies to preserve student agency and calibrate trust.

ELEVATE supports agency-preserving interaction by design: (i) it can default to coaching behaviors (eliciting reasoning and partial attempts) rather than direct solution delivery; (ii) it can expose transparency cues (uncertainty markers, verification prompts, and—when enabled—document-grounding signals); and (iii) it can provide progressive scaffolding (goal $\rightarrow$ plan $\rightarrow$ attempt $\rightarrow$ feedback $\rightarrow$ reflection). This aligns with process-oriented learning models and helps maintain a meaningful distinction between assistance and substitution. Moreover, because students may engage the system for more than academic questions, institutions must define boundaries for wellbeing-related dialogue and escalation procedures for high-risk content. Local-first deployment improves data protection but does not eliminate safeguarding obligations; instead, it increases the institution’s responsibility to configure and audit safety policies over time, including defining what the tutor should refuse, redirect, or escalate.

\subsubsection{Teacher Role and Pedagogy}
\label{subsec:teacher_role}

An embodied LLM-tutor is likely to be used differently than a conventional chatbot: spoken dialogue and animated feedback encourage short, frequent exchanges that can support micro-scaffolding (stepwise hints, checks for understanding, recap prompts), but may also compete with teacher-led pacing if the system becomes a parallel instructional channel. For this reason, ELEVATE should be framed as a pedagogical instrument integrated into lesson plans (e.g., warm-up, guided practice, revision stations) rather than as a general-purpose substitute for instruction. Because embodiment and voice can amplify authority effects, the avatar should adopt a careful epistemic stance (signaling uncertainty, encouraging verification, and modeling good reasoning practices; e.g., inviting students to consult notes or textbooks). Personalization can further support differentiation (vocabulary, pacing, level of guidance), but it should remain transparent and teacher-governed; a practical approach is to provide explicit classroom modes (e.g., primary school, special needs support, exam preparation) selected by teachers rather than inferred silently. Finally, to align with school norms and reduce pedagogical risk, ELEVATE should expose the following teacher-in-the-loop control points: (i) content boundaries (age-appropriate constraints, curriculum alignment), (ii) response policies (hint-first behaviors, refusals, optional source-grounding), (iii) override and escalation (pause/hand-off to teacher for sensitive cases), and (iv) classroom norms (interaction time limits and prompts that encourage collaborative discussion).

\subsection{Authorship and Integrity}
\label{subsec:integrity}

Generative tools challenge authorship norms across writing, coding, and take-home tasks, so adoption should be paired with explicit, actionable usage rules (e.g., allowed: brainstorming, outlining, hints; disallowed: submitting generated final answers without attribution). 

The avatar interface can reinforce these norms by prompting attribution and prioritizing ``learning-first'' modes such as hints, critique, and rubric-based feedback, which reduce incentives for copying. More broadly, if students can access tutoring on demand, assessment should shift toward evaluating understanding and process (oral defenses, in-class problem solving, reflective journaling, and stepwise submissions that capture reasoning).
Moreover, local-first architectures like ours can support this shift by enabling privacy-preserving institutional oversight and pedagogical analytics, such as identifying recurring misconceptions, estimating time-on-task patterns, or summarizing which concepts required repeated scaffolding.

However, these benefits must be balanced against proportionality, student trust, and data protection obligations. A pragmatic safeguard is to default to aggregate and minimal logging (e.g., interaction metadata, policy/configuration versions, and high-level summary signals) while restricting retention of raw content (full transcripts or audio) to clearly justified cases with defined governance and limited access.

\subsection{System Assessment and Control}
\label{subsec:Assessment}

To operationalize integrity-aware adoption within our platform, we discuss two complementary assessment tracks: (i) teacher-facing evaluation of governance, alignment, and sustainability, and (ii) student-facing evaluation of learning evidence, process transparency, and responsible use.

\subsubsection{Teacher-Facing Assessment: Governance, Pedagogical Alignment, and Operational Control}
\label{subsubsec:teacherAssessment}

From the institutional perspective, assessment must verify that the platform remains governable, curricularly aligned, and operationally sustainable over time. This entails evaluating not only the quality of generated outputs, but also how content is configured, constrained, reviewed, and integrated into teaching practice. First, the platform should support configuration and policy audits: teachers (and, where applicable, school leadership) should be able to inspect which prompt templates, refusal rules, scope constraints, and grounding sources were active for a given classroom or assignment, and whether those settings were updated consistently across time. The assessment should therefore track configuration revisions, policy bundle versions, and the frequency and rationale of changes (e.g., curriculum updates, emerging misuse patterns, or safeguarding adjustments), enabling accountability and reproducibility of classroom behavior.

Second, teacher-facing assessment should measure pedagogical alignment and instructional utility. Rather than treating the tutor as a general conversational agent, educators should evaluate whether its responses consistently implement the intended instructional strategy (e.g., hint-first scaffolding, Socratic prompting, rubric-based critique) and whether this strategy matches course-level learning objectives. This can be operationalized through periodic sampling of interactions (or summaries) using teacher-defined rubrics that check: correctness relative to the taught curriculum, appropriateness of difficulty, coherence with assessment rules (e.g., refusing to provide final answers for restricted tasks), and quality of formative feedback. Importantly, alignment checks should include failure modes (hallucinations, overconfident tone, non-compliance with constraints) and document how frequently they occur under normal use.

Third, the assessment must capture teacher workload and adoption friction. Teachers should evaluate the time and cognitive effort required to configure lessons, curate grounding materials, review summaries, and handle escalation events. In practice, the platform should surface indicators such as: the time spent per configuration cycle, the frequency of manual corrections or overrides, the rate of policy-triggered refusals that require teacher follow-up, and the perceived usability of governance controls. These measures are essential because governance mechanisms that are theoretically robust but operationally heavy are unlikely to be sustained in real classrooms.

Finally, teacher-facing assessment should include privacy, proportionality, and accountability checks. Teachers (and institutions) should periodically review whether logging and data retention remain proportionate to pedagogical aims, whether content summaries are sufficient without exposing sensitive student data, and whether access control to logs is appropriately limited. This ensures that institutional oversight does not drift into surveillance and that auditing remains compatible with trust and safeguarding obligations.

\subsubsection{Student-Facing Assessment: Learning Evidence, Process Transparency, and Responsible Use}
\label{subsubsec:studentAssessment}

For students, assessment should be designed to produce observable evidence of understanding in an environment where the platform can provide on-demand guidance. Consequently, evaluation must focus on how content is produced and used, not merely on the final artifact. 
A first axis is process transparency: students should be assessed on their ability to document and justify how platform assistance contributed to their work. This can be implemented through short structured reflections (e.g., ``what I asked'', ``what I changed'', ``what I learned''), assistance summaries attached to submissions, and checkpoints requiring students to explain why specific revisions were made. The goal is to normalize disclosure and make responsible tool use part of academic practice rather than an exception.

A second axis concerns reasoning and transfer. Since the platform can generate fluent explanations, assessments should emphasize tasks where students must demonstrate comprehension through application: in-class problem solving, oral defenses, step-by-step derivations, code walk-throughs, debugging sessions, or micro-vivas where students explain choices and respond to variations. These formats directly test whether learners can reconstruct the reasoning behind an output and transfer it to new instances, reducing the likelihood that performance is driven by passive copying.

A third axis is metacognitive and self-regulatory competence. Because generative tutors can encourage over-reliance, assessment should explicitly value behaviors such as goal setting, error diagnosis, and iterative improvement. Platforms like ours can support this by encouraging hint-first interactions and by enabling structured self-assessment prompts (e.g., asking students to predict an answer before requesting help, or to rate confidence and identify uncertainties). Student evaluation can therefore include evidence of revision quality, responsiveness to feedback, and the capacity to identify misconceptions and correct them over time.

Finally, student-facing assessment should monitor safety and inclusion outcomes. In heterogeneous classrooms, students differ in accessibility needs, language proficiency, and susceptibility to authority effects. Assessment should therefore include measures of perceived clarity, cognitive load, and accessibility effectiveness (e.g., whether multimodal options reduce barriers), alongside indicators of trust calibration (e.g., whether students cross-check outputs, ask for justifications, and recognize uncertainty). These measures help ensure that the platform supports equitable learning rather than amplifying existing disparities in confidence, access, or support.

\subsection{Institutional Adoption and Accessibility}
\label{subsec:governance}

A core premise of this work is that GenAI adoption in education is not only a technical integration problem, but a socio-technical one: the same system architecture must remain pedagogically useful while adapting to heterogeneous institutional conditions (e.g., infrastructure, staff expertise, student population size) and to non-negotiable ethical requirements (e.g., privacy-by-design, accessibility). For this reason, ELEVATE was designed as a configurable three-stratum loop, where the separation between (i) student interaction, (ii) teacher governance, and (iii) local GenAI execution enables multiple deployment profiles without changing the pedagogical paradigm.

\subsubsection{Rationale for a configuration space}
Schools differ substantially in (a) \textbf{resource richness} (availability of devices, network stability, IT support), (b) \textbf{number of students} (which impacts concurrency and operational load), and (c) \textbf{school type} (primary/secondary/higher education or special education), which determines interaction modalities, accessibility needs, and curricular constraints. A single fixed architecture tends to privilege well-resourced contexts and produces unequal adoption opportunities. ELEVATE instead frames deployment as a \textbf{controlled configuration space} that allows institutions to trade off interaction richness and throughput while preserving ethical invariants.

\subsubsection{Ethicality as invariants rather than options}
We treat ethicality as a set of \emph{invariants} that must hold across all configurations:
(i) \textbf{privacy-by-design} (local/on-prem processing; minimal data retention),
(ii) \textbf{accessibility-by-default} (captions, alternative inputs, reading-level controls),
(iii) \textbf{institutional governance} (teacher-defined scope, document control, and refusal policies).
This design turns ethics from a post-hoc compliance layer into a first-class systems requirement, consistent with European regulatory and cultural expectations around educational technology.
Moreover, by making deployment explicitly configurable along these axes while keeping ethicality invariant, ELEVATE aims to reduce structural barriers to adoption and mitigate the risk that GenAI-enhanced education becomes available only to well-funded institutions. In this sense, the architecture is not merely efficient: it is designed to support a pluralistic educational landscape, where heterogeneous schools can implement comparable pedagogical capabilities under different constraints, and where governance remains anchored in educators rather than implicit model behavior.

\subsubsection{Mapping institutional variables to system configurations}
We therefore justify the key architectural choices (modularity, local-first execution, and teacher-governed constraints) as enabling three practical configuration axes:

\begin{enumerate}
    \item \textbf{Resource richness (low $\rightarrow$ high).}
    In low-resource schools, the local-first design supports offline operation and avoids recurring costs (cloud subscriptions, bandwidth). Here, ELEVATE can employ smaller quantized models, limited multimodal rendering, and lightweight avatar assets while maintaining accessibility features (e.g., captions, simplified language). In high-resource contexts, the same framework can increase avatar fidelity, enable richer multimodal sensing, and support larger on-prem models or dedicated servers, without changing the governance structure.

    \item \textbf{Number of students (small $\rightarrow$ large).}
    Student population size impacts scalability requirements. For small cohorts, execution can be fully on-device or on a single shared workstation. For large cohorts, the modular engine may support on-prem pooling (e.g., one or more local servers) and asynchronous orchestration (streaming output, queue-based scheduling) to maintain responsiveness under concurrency. Importantly, scaling is achieved without exporting student data to external services, preserving ethical invariants.

    \item \textbf{School type (developmental stage and pedagogical regime).}
    Different school types require different interaction and assessment constraints: primary education may prioritize short utterances, simplified language, and strong turn-taking cues; secondary education may require structured explanations, quizzes, and rubric-based feedback; vocational education may require document-grounded procedural guidance; special education contexts may require AAC-compatible phrasing and stronger accessibility defaults. ELEVATE operationalizes these differences at the Teacher Governance Layer, converting them into enforceable constraints (prompting rules, retrieval scope, output formatting, refusal policies), while the Student Layer adapts the avatar interface to age-appropriate and accessible modalities.
\end{enumerate}

\subsection{Governance, Accountability, and Inclusion}
\label{subsec:equity}

Local-first deployment pushes towards a shift in operational ownership: schools become responsible for updates, safety configuration, device management, and policy alignment, which requires a realistic maintenance model (district IT, a designated AI coordinator, or an audited service provider operating under school control).
Staff training is similarly essential, focusing on classroom integration, limitations of model outputs, and scenario-based responses to misinformation, biased content, or sensitive prompts. Accountability must be clearly defined: when the system produces harmful or incorrect outputs, institutions need incident response procedures, rapid policy/template update mechanisms, and version tracking; governance should include periodic audits (content sampling, bias checks, and usability checks) and documentation of configuration changes to avoid drift across classrooms and schools.

Considering equity and inclusion, ELEVATE has lower hardware requirements and the potential to support local languages (changing the LLM) and curricula without continuous connectivity (with additional prompting engineering). However, inclusion depends on cultural and linguistic alignment of the avatar's persona and examples, and on adaptable accessibility modalities (speech, captions, simplified language, adjustable pace, and the ability to switch between text-only, voice-only, or combined interaction). At the same time, embodiment can create barriers for some learners (e.g., sensory sensitivities or discomfort with anthropomorphic agents), also making its configuration a strong requirement.

\section{Conclusion and Future Work}



This work presented ELEVATE, a framework-oriented perspective on the design and deployment of LLM-driven virtual teachers for educational contexts, emphasizing ethical adoption, accessibility, and pedagogical intent. By abstracting architectural principles from specific implementations, the proposed approach highlights how generative systems can support learning without reducing education to mere content delivery. The ELEVATE framework and its reference implementation illustrate how real-time and multimodal interaction (through coordinated dialogue, speech, and avatar embodiment) may contribute to meaningful understanding while remaining adaptable to diverse institutional and technological constraints.

Future work may extend the proposed framework by analyzing domain-specific and user-centered scenarios in greater depth. In particular, targeted studies could explore the use of virtual teachers as adaptive support tools for students with learning difficulties, as assistive companions for senior learners, or as mediators for international students facing linguistic and cultural barriers. From a technical perspective, we plan to implement automatic document-driven fine-tuning mechanisms (and optionally RAG) as first-class customization strategies. This would enable institutions to specialize the tutor on tightly scoped educational use cases (e.g., a specific course module, school policy, or local curriculum) by grounding responses in approved corpora and, when feasible, adapting model behavior to domain terminology and instructional style. Such extensions would further strengthen transparency and governance by linking tutor outputs to explicitly curated knowledge sources, while preserving the local-first and privacy-by-design principles that motivate ELEVATE.
Finally, we will apply controlled user studies in real school settings (where we deployed our setting), evaluating both learning outcomes and socio-technical effects.

\section*{Declarations}
\paragraph{Declaration of generative AI and AI-assisted technologies in the manuscript preparation process.} 
The authors declare that we adopted ChatGPT 5.2 to paraphrase selected sentences for clarity and paragraph restructuring; all scientific claims, interpretations, and final wording decisions remain the responsibility of the authors.

\paragraph{Declaration of competing interests}
The authors declare that they have no known competing financial interests or personal relationships that could have appeared to influence the work reported in this paper.

\paragraph{Funding statement}
This research did not receive any specific grant from funding agencies in the public, commercial, or not-for-profit sectors


\bibliographystyle{elsarticle-harv} 
\biboptions{round,comma}

\bibliography{cas-refs}

\end{document}